\documentclass[aps,twocolumn,showpacs,footinbib]{revtex4}%
\usepackage[latin9]{inputenc}
\usepackage{amssymb}
\usepackage{amsmath}
\usepackage{amscd}
\usepackage{graphicx}
\usepackage{subfigure}
\usepackage{float}
\begin{document}
\draft
\title{Virtual-photon-induced quantum phase gates for two distant atoms trapped in
separate cavities}
\author{Shi-Biao Zheng\thanks{%
E-mail: sbzheng@pub5.fz.fj.cn}}
\address{Department of Physics\\
Fuzhou University\\
Fuzhou 350002, P. R. China}
\date{\today }

\begin{abstract}
We propose a scheme for implementing quantum gates for two atoms trapped in
distant cavities connected by an optical fiber. The effective long-distance
coupling between the two distributed qubits is achieved without excitation
and transportation of photons through the optical fiber. Since the cavity
modes and fiber mode are never populated and the atoms undergo no
transitions the gate operation is insensitive to the decoherence effect when
the thermal photons in the environment are negligible. The scheme opens
promising perspectives for networking quantum information processors and
implementing distributed and scalable quantum computation.
\end{abstract}

\pacs{PACS number: 03.67.Lx, 03.67.Mn, 42.50.Dv}

\vskip 0.5cm \maketitle \narrowtext

In recent years, much attention has been paid to quantum computers, which
are based on the superposition principle of quantum mechanics. This type of
new machines can solve some problems much more efficient than classical
computers. Typical examples are the factorization of a large composite
number via Shor's algorithm [1] and the search for an item from a disordered
system via Grover's algorithm [2]. It has been shown that the building
blocks of quantum computers are two-quantum-bit (qubit) logic gates [3].
Such gates have been demonstrated in cavity QED [4], ion trap [5], and NMR
[6] experiments. The cavity QED system with atoms interacting with quantized
electromagnetic fields is almost an ideal candidate for implementing quantum
information processors because atoms are suitable for storing information
and photons suitable for transporting information. A number of robust
schemes have been proposed for implementing quantum gates for two atoms
trapped in a cavity [7-9].

The implementation of a quantum computational task corresponds to the
performance of an unitary transformation on the quantum register, which is
composed of multiple qubits [10]. For building a quantum computer that has
practical applications a large number of qubits should be ensembled and the
power of quantum computers increases as the number of qubits increases.
However, there are physical limitations on the number of qubits in a quantum
computer and large-scale quantum computation has not been experimentally
achieved yet. To overcome this limit, several distinct quantum information
processors should be connected via quantum channels to form a powerful
quantum computer. Entanglement and controlled-phase gates between two
distant atoms have been proposed [11,12]. The schemes are probabilistic and
the success probability depends upon the efficiency of photodetectors.
Moreover, the success probability decreases as the number of qubits
increases.

Schemes have been proposed for quantum communication between two atoms
trapped in distant optical cavities connected by an optical fiber [13].
These schemes are based on accurately tailored sequences of pulses or
adiabatic passage. Recently, Serafini et al. [14] suggested a scheme for
realizing quantum gates between two two-level atoms in distant optical
cavities coupled by an optical fiber. The scheme is based on the Rabi
oscillation of the whole system composed of the atoms, cavity modes, and
fiber mode. Yin and Li generalized the idea to the multi-atom case [15]. In
all of the previous schemes for quantum communication the quantum state is
mediated by photons carried by the optical fiber and thus the fidelity is
significantly affected by the loss of photons. Furthermore, in Refs. [14]
and [15] the controlled-phase gates are obtained under the condition that
the atom-cavity coupling strength is much smaller than the cavity-fiber
coupling, which might require the weak couplings. In the weak coupling
regime, the decoherence effect arising from the atomic spontaneous emission
and cavity decay would invalid the scheme. It has been shown that a quantum
computer can work only when the individual gate infidelity is below a
certain constant threshold, which is about $10^{-2}$ [16]. The
implementation of deterministic high-fidelity quantum gates between two
qubits at different nodes is a prerequisite for realizing distributed and
scalable quantum computation.

In this paper we propose an alternative scheme for controlled-phase gates
between two atoms trapped in two distant optical cavities. The distinct
feature of our scheme is that the two distant qubits are coupled without
exciting and transferring photons through the optical fiber, which is in
contrast with the previous quantum networking schemes. The long-range
interaction is mediated by the vacumm fields of the cavities and fiber and
the whole system evolves in the decoherence-free subspace, in which neither
of the subsystems is excited. Thus the scheme is insensitive to the atomic
spontaneous emission, cavity decay, and fiber loss when the thermal photons
in the environment are negligible. Furthermore, our scheme does not require
the atom-cavity coupling to be smaller than the cavity-fiber coupling. These
features make the scheme very promising for implementation of distributed
and scalable quantum computation that can avoid decoherence.

We consider that two identical atoms are trapped in distant cavities
connected by a single-transverse-mode optical fiber, as shown in Fig. 1. The
number of longitudinal modes of the fiber that significantly interact with
the corresponding cavity modes is on the order of $l\stackrel{-}{\nu }/(2\pi
c)$, where $l$ is the length of the fiber and $\stackrel{-}{\nu }$ is the
decay rate of the cavities' fields into the continuum of the fiber modes. In
the short fiber limit $l\stackrel{-}{\nu }/(2\pi c)\leq 1$, only one fiber
mode essentially interacts with the cavity modes [14]. In this case the
coupling between the cavity modes and fiber is given by the interaction
Hamiltonian
\begin{equation}
H_{c,f}=\nu b(a_1^{\dagger }+e^{i\varphi }a_2^{\dagger })+H.c.,
\end{equation}
where $b$ is the annihilation operator for the fiber mode, $a_j^{\dagger }$
is the creation operator for the $j$th cavity mode, $\nu $ is the
cavity-fiber coupling strength, and $\varphi $ is the phase due to
propagation of the field through the fiber. Each atom has one excited state $%
\left| e\right\rangle $ and two ground states $\left| g\right\rangle $ and $%
\left| f\right\rangle $, as shown in Fig. 2. The transition $\left|
e\right\rangle \longleftrightarrow \left| g\right\rangle $ is coupled to the
corresponding cavity mode with the coupling constant $g$ and a classical
laser field with the Rabi frequency $\Omega $. The detunings of the cavity
mode and classical field are $\Delta $ and $\Delta -\delta $, respectively.
During the interaction, the state $\left| f\right\rangle $ is not affected.
In the interaction picture, the Hamiltonian describing the atom-field
interaction is
\begin{equation}
H_{a,c}=\sum_{j=1}^2[(ga_je^{i\Delta t}+\Omega e^{i(\Delta -\delta
)t})\left| e_j\right\rangle \left\langle g_j\right| +H.c.].
\end{equation}
Define $c_0=\frac 1{\sqrt{2}}(a_1-e^{i\varphi }a_2)$, $c_1=\frac 12%
(a_1+e^{i\varphi }a_2+\sqrt{2}b)$, and $c_2=\frac 12(a_1+e^{i\varphi }a_2-%
\sqrt{2}b)$. Here $c_0$, $c_1$, and $c_2$ are three bosonic modes, which are
linearily relative to the field modes of the cavities and fiber. Then we can
rewrite the whole Hamiltonian in the interaction picture as
\begin{equation}
H=H_0+H_i,
\end{equation}
where
\begin{equation}
H_0=\sqrt{2}\nu c_1^{\dagger }c_1-\sqrt{2}\nu c_2^{\dagger }c_2,
\end{equation}
and
\begin{eqnarray}
H_i &=&[\frac 12g(c_1+c_2+\sqrt{2}c_0)e^{i\Delta t}+\Omega
e^{i(\Delta -\delta )t}]\left| e_1\right\rangle \left\langle
g_1\right|\cr &&\ \ \ +[\frac 12g(c_1+c_2-\sqrt{2}c_0)e^{-i\varphi
}e^{i\Delta t}\cr&&+\Omega e^{i(\Delta -\delta )t}]\left|
e_2\right\rangle \left\langle g_2\right| +H.c.
\end{eqnarray}

We now perform the unitary transformation $e^{iH_0t}$, and obtain
\begin{eqnarray}
H_i^{^{\prime }} &=&\{\frac 12g[e^{i(\Delta -\sqrt{2}\nu )t}c_1+e^{i(\Delta +%
\sqrt{2}\nu )t}c_2+\sqrt{2}e^{i\Delta t}c_0]\cr&&+\Omega e^{i(\Delta
-\delta )t}\}\left| e_1\right\rangle \left\langle g_1\right|
+\{\frac 12ge^{-i\varphi }[e^{i(\Delta -\sqrt{2}\nu
)t}c_1\cr&&+e^{i(\Delta +\sqrt{2}\nu )t}c_2-\sqrt{2}e^{i\Delta
t}c_0]+\Omega e^{i(\Delta -\delta )t}\}\left| e_2\right\rangle
\left\langle g_2\right|\cr&& +H.c..
\end{eqnarray}
We here assume that $\Delta \gg \sqrt{2}\nu $, $\delta $, $g$, $\Omega $.
Then the atoms do not exchange energy with the cavity modes, fiber mode, and
classical fields due to the large detuning. The quantum information is
encoded in the two ground states $\left| g\right\rangle $ and $\left|
f\right\rangle $. Since both the atoms are initially populated in the ground
states, they can not exchange excitation with each other via the virtual
excitation of the cavity modes and they remain in the ground states.
However, the three bosonic modes $c_0$, $c_1$, and $c_2$ can be coupled to
each other and the classical fields via the virtual excitation of the atoms.
So the Hamiltonian $H_i^{^{\prime }}$ can be replaced by
\begin{eqnarray}
H_i^{^{\prime \prime }} =-[\lambda _1e^{i(\delta -\sqrt{2}\nu
)t}c_1+\lambda _2e^{i(\delta +\sqrt{2}\nu )t}c_2+\lambda
_0e^{i\delta t}c_0]\left| g_1\right\rangle \left\langle g_1\right|
\cr
\ -[\lambda _1e^{i(\delta -\sqrt{2}\nu )t}c_1+\lambda _2e^{i(\delta +\sqrt{%
2}\nu )t}c_2-\lambda _0e^{i\delta t}c_0]e^{-i\varphi }\left|
g_2\right\rangle \cr \left\langle g_2\right| -(\xi
_1e^{-2i\sqrt{2}\nu t}c_1c_2^{\dagger }+\xi _2e^{-i\sqrt{2}\nu
t}c_1c_0^{\dagger }\cr+\xi _0e^{-i\sqrt{2}\nu t}c_0c_2^{\dagger
})\left| g_1\right\rangle \left\langle g_1\right|   -(\xi
_1e^{-2i\sqrt{2}\nu t}c_1c_2^{\dagger }\cr-\xi _2e^{-i\sqrt{2}\nu
t}c_1c_0^{\dagger }-\xi _0e^{-i\sqrt{2}\nu t}c_0c_2^{\dagger
})\left|
g_2\right\rangle \left\langle g_2\right| +H.c.  \nonumber \\
-(\eta +\varepsilon _1c_1^{\dagger }c_1+\varepsilon _2c_2^{\dagger
}c_2+\varepsilon _0c_0^{\dagger }c_0)(\left| g_1\right\rangle
\left\langle g_1\right| +\left| g_2\right\rangle \left\langle
g_2\right| ),  \cr
\end{eqnarray}
where $\lambda _0=\frac{\sqrt{2}g\Omega }4(\frac 1\Delta +\frac 1{\Delta
-\delta })$, $\lambda _1=\frac{g\Omega }4(\frac 1{\Delta -\sqrt{2}\nu }+%
\frac 1{\Delta -\delta })$, $\lambda _2=\frac{g\Omega }4(\frac 1{\Delta +%
\sqrt{2}\nu }+\frac 1{\Delta -\delta })$, $\xi _1=\frac{g^2}4(\frac 1{\Delta
+\sqrt{2}\nu }+\frac 1{\Delta +\sqrt{2}\nu })$, $\xi _2=\frac{\sqrt{2}g^2}4(%
\frac 1{\Delta -\sqrt{2}\nu }+\frac 1\Delta )$, $\xi _3=\frac{\sqrt{2}g^2}4(%
\frac 1{\Delta +\sqrt{2}\nu }+\frac 1\Delta )$, $\eta =\frac{\Omega ^2}{%
\Delta -\delta }$, $\varepsilon _0=\frac{g^2}{4\Delta }$, $\varepsilon _1=%
\frac{g^2}{4(\Delta -\sqrt{2}\nu )}$, and $\varepsilon _1=\frac{g^2}{%
4(\Delta +\sqrt{2}\nu )}$. Here $\lambda _j$ ($j=0,1,2$) is the effective
coupling between the bosonic mode $c_j$ and the classical fields, $\xi _j$
is the mode-mode coupling, and $\eta $ and $\varepsilon _j$ describe the
Stark shifts induced by the classical fields and the bosonic mode $c_j$,
respectively.

Under the conditions $\delta $, $\sqrt{2}\nu $, $\delta -\sqrt{2}\nu $, $%
\delta +\sqrt{2}\nu \gg \lambda _0$, $\lambda _1$, $\lambda _2$, $\xi _0$, $%
\xi _1$, $\xi _2$, the bosonic modes $c_0$, $c_1$, and $c_2$ and can not
exchange energy with each other and with the classical fields. The
nonresonant couplings between the bosonic modes and the classical fields
lead to energy shifts depending upon the number of atoms in the state $%
\left| g\right\rangle $. Meanwhile, the nonresonant mode-mode couplings
cause energy shifts depending upon both the excitation numbers of the modes
and the number of atoms in the state $\left| g\right\rangle $. The effective
Hamiltonian is
\begin{eqnarray}
H_e &=&(\mu _1+\mu _2)(\left| g_1\right\rangle \left\langle
g_1\right| +\left| g_2\right\rangle \left\langle g_2\right| )^2\
\cr&&+\mu _0(\left| g_1\right\rangle \left\langle g_1\right| -\left|
g_2\right\rangle \left\langle g_2\right| )^2 +\frac{\xi
_1^2}{2\sqrt{2}\nu }(c_1c_1^{\dagger }-c_2c_2^{\dagger
})\cr&&(\left| g_1\right\rangle \left\langle g_1\right| +\left|
g_2\right\rangle \left\langle g_2\right| )^2
+[\frac{\xi _2^2}{\sqrt{2}\nu }(c_1c_1^{\dagger }-c_0c_0^{\dagger })\cr&&+\frac{%
\xi _0^2}{\sqrt{2}\nu }(c_0c_0^{\dagger }-c_2c_2^{\dagger })](\left|
g_1\right\rangle \left\langle g_1\right| -\left| g_2\right\rangle
\left\langle g_2\right| )^2 \cr&&-(\eta +\varepsilon _1c_1^{\dagger
}c_1+\varepsilon _2c_2^{\dagger }c_2+\varepsilon _0c_0^{\dagger
}c_0)(\left| g_1\right\rangle \left\langle g_1\right| +\left|
g_2\right\rangle \left\langle g_2\right| ),  \cr&&
\end{eqnarray}
where $\mu _0=\frac{\lambda _0^2}\delta $, $\mu _1=\frac{\lambda _1^2}{%
\delta -\sqrt{2}\nu }$, and $\mu _2=\frac{\lambda _2^2}{\delta +\sqrt{2}\nu }
$. Here $\mu _j$ ($j=0,1,2$) is the effective coupling between the atoms due
to the nonresonant coupling between the bosonic mode $c_j$ and the classical
fields. The Hamiltonian describes a four-photon process which is induced by
the virtual excitation of the atoms and bosonic modes. The quantum numbers
of the bosonic modes $c_0$, $c_1$, and $c_2$ conserve during the
interaction. Suppose that the two cavity modes and the fiber mode are all
initially in the vacuum state (at the optical frequencies, the thermal
photons are negligible). Then the three bosonic modes $c_0$, $c_1$, and $c_2$
remain in the vacuum state during the evolution. In this case the effective
Hamiltonian reduces to
\begin{eqnarray}
H_e &=&(\mu _1+\mu _2)(\left| g_1\right\rangle \left\langle
g_1\right| +\left| g_2\right\rangle \left\langle g_2\right| )^2\
\cr&&+\mu _0(\left| g_1\right\rangle \left\langle g_1\right| -\left|
g_2\right\rangle \left\langle g_2\right| )^2 \cr &&\ -\eta (\left|
g_1\right\rangle \left\langle g_1\right| +\left| g_2\right\rangle
\left\langle g_2\right| ).
\end{eqnarray}
Due to the nonresonant coupling between the classical fields and vacuum
bosonic modes induced by the virtual excitation of the atoms, the atomic
system undergoes an energy shift, which is nonlinear in the number of the
atoms in the ground state $\left| g\right\rangle $. The nonlinear energy
shift leads to the evolution
\begin{equation}
\begin{array}{c}
\left| g_1\right\rangle \left| g_2\right\rangle \longrightarrow e^{-i(4\mu
_1+4\mu _2-2\eta )t}\left| g_1\right\rangle \left| g_2\right\rangle ,\quad
\\
\left| g_1\right\rangle \left| f_2\right\rangle \longrightarrow e^{-i(\mu
_1+\mu _2+\mu _0-\eta )t}\left| g_1\right\rangle \left| f_2\right\rangle ,
\\
\left| f_1\right\rangle \left| g_2\right\rangle \longrightarrow e^{-i(\mu
_1+\mu _2+\mu _0-\eta )t}\left| f_1\right\rangle \left| g_2\right\rangle ,
\\
\left| f_1\right\rangle \left| f_2\right\rangle \longrightarrow \left|
f_1\right\rangle \left| f_2\right\rangle .
\end{array}
\end{equation}
A two-qubit quantum phase gate is obtained after the qubit-qubit coupling
and the single-qubit phase shifts: $\left| g_j\right\rangle \longrightarrow
e^{i(\mu _1+\mu _2+\mu _0-\eta )t}\left| g_j\right\rangle $. The atomic
system undergoes a conditional phase shift $-2(\mu _1+\mu _2-\mu _0)t$ if
and only if the two atoms are initially in the state $\left|
g_1\right\rangle \left| g_2\right\rangle $. The conditional phase shift is
adjustable via the interation time. For the implementation of a given
quantum computational task, the single-qubit phase shifts might be
unnecessary as they can be absorbed into the next single-qubit rotations.
During the operation, all the the atoms, cavity modes and fiber mode are not
excited and the decoherence is suppressed when the thermal photons in the
environment are negligible.

We briefly address the experimental feasibility of the proposed scheme. Set $%
\Omega =g$, $\Delta =30g$, $\delta =g$, $\nu =\sqrt{2}g$, and $\Gamma
=\kappa =0.01g$, where $\Gamma $ and $\kappa $ are the decay rates for the
atomic excited state and the bosonic modes, respectively. In this case the
probability that the atoms undergo a transition to the excited state due to
the off-resonant interaction with the classical fields is $P_1\simeq \Omega
^2/\Delta ^2\simeq 1.11\times 10^{-3}$. Meanwhile, the probability that the
three modes $c_0$, $c_1$, and $c_2$ are excited due to nonresonant coupling
with the classical modes is $P_2\simeq \lambda _0^2/\delta ^2+\frac{\lambda
_1^2}{(\delta -\sqrt{2}\nu )^2}+\frac{\lambda _2^2}{(\delta +\sqrt{2}\nu )^2}%
\simeq 0.917\times 10^{-3}$. Therefore, the effective Hamiltonian (11) and
(13) is valid. During the procedure all the atoms and bosonic modes are only
virtually excited. The effective decoherence rates due to the atomic
spontaneous emission and the decay of the bosonic modes are $\Gamma
^{^{\prime }}\simeq P_1\Gamma \simeq $ $1.11\times 10^{-5}g$ and $\kappa
^{^{\prime }}\simeq P_2\kappa \simeq 0.917\times 10^{-5}g$, respectively.
The time needed to produce a conditional phase $0.15\pi $ is $t=101.25\pi /g$%
. The infidelity induced by the decoherence is about $(\Gamma ^{^{\prime
}}+\kappa ^{^{\prime }})t\simeq 0.645\times 10^{-2}$, which is below the
threshold $0.01$ for fault-tolerant computing and error correction. For the
same decoherence rates $\Gamma =\kappa =0.01g$, the infidelity of such a
gate achieved in the scheme of Ref. [14] is about $0.7\times 10^{-1}$. The
infidelity increases with the conditional phase. For $\Gamma =\kappa =0.01g$
the infidelity of the $\pi $-phase gate in our scheme is about $4.3\times
10^{-2}$. In contrast, the decoherence completely deteriorates the fidelity
of this gate in Ref. [14], which is produced after six Rabi oscillations. It
should be noted that the scheme of Ref. [14] requires that the atom-cavity
coupling strength be much smaller than the cavity-fiber coupling for the
implementation of the controlled-phase gates. In this case it is difficult
to achieve strong atom-cavity coupling. In order to estimate the length of
the fiber, let us set $\stackrel{-}{\nu }=1$ GHz [14]. Thus we have $l\leq
1.884$ m. In this case the round-trip time of virtual photons is $t_r\sim
10^{-8}$ s. For $g=2\pi \times 34$ MHz [17], the operation time scale is on
the order of $10^{-6}$, much longer than $t_r$. Thus the interaction
Hamiltonian holds.

For simplicity, we have assumed that the two atoms have the same coupling
with the corresponding cavity modes. Suppose that the two couplings are $%
g_1=g$ and $g_2=rg$, respectively. Then the conditional phase is $-2r(\mu
_1+\mu _2-\mu _0)t$. Unlike the scheme of Ref. [14], the conditional phase
gate is valid no matter when the two couplings are the same or not. Another
advantage of the gate is that the dynamics can be easily controlled. After
the gate operation, the interaction can be frozen by switching off the
classical fields. In a recent experiment [18], the localization to the
Lamb-Dicke limit of the axial motion was demonstrated for a single atom
trapped in an optical cavity. The atomic storage time is on the order of 1
s, much longer than the operation time. The near perfect fiber-cavity
coupling with the efficiency larger than 99.9\% can be realized using
fiber-taper coupling to high-Q silica microspheres [19].

In conclusion, we have described a protocol for implementing conditional
phase gates for two atoms trapped in separate cavities conneced by an
optical fiber. Our scheme does not involve the excitation and transportation
of photons. The long-distance qubit-qubit interaction is obtained via
virtual excitation of the atoms, cavity modes, and fiber mode.Thus the
scheme is insensitive to the atomic spontaneous emission, cavity decay, and
fiber loss when the thermal photons in the environment are negligible. It is
unnecessary to require that the atom-cavity coupling be smaller than the
cavity-fiber coupling. The idea may be straightly extended to generate
cluster states for multiple atoms each trapped in an optical cavity, which
are the resources for the one-way quantum computation [20]. The scheme opens
promising perspectives for the realization of distributed and scalable
quantum network.

This work was supported by the National Natural Science Foundation of China
under Grant No. 10674025 and the Doctoral Foundation of the Ministry of
Education of China under Grant No. 20070386002.

\begin{figure}
\centering
\includegraphics[width=0.2\columnwidth,angle=90]{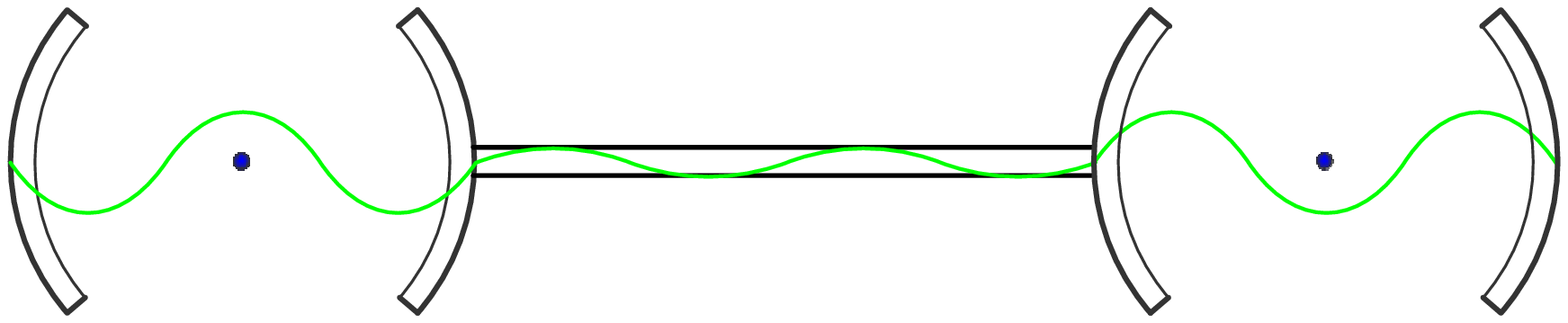} \caption{(Color
online)  The experimental setup. Two distant atoms are trapped in
separate cavities connected by an optical fiber. The two cavity
modes are coupled to the fiber mode with the coupling strength $\nu
$.}
\end{figure}

\begin{figure}
\centering
\includegraphics[width=1\columnwidth,angle=90]{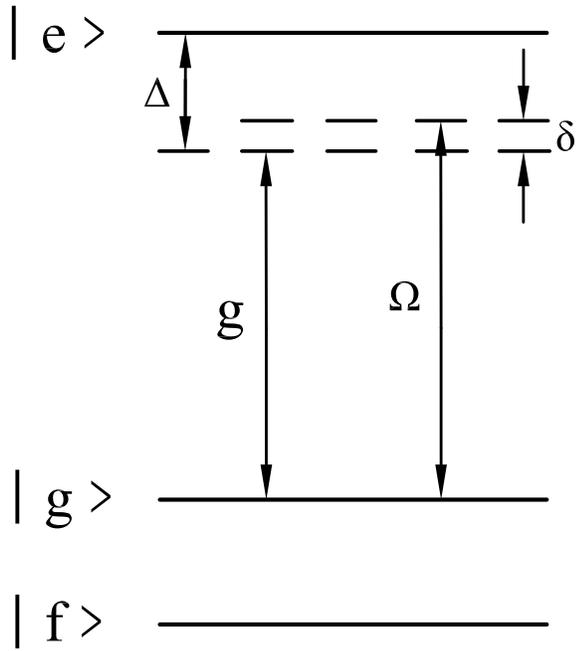} \caption{(Color
online)  The atomic level configuration. The transition $\left|
e\right\rangle \longleftrightarrow \left| g\right\rangle $ is
coupled to the corresponding cavity mode with the coupling constant
$g$ and a classical laser field with the Rabi frequency $\Omega $.
The detunings of the cavity mode and classical field are $\Delta $
and $\Delta -\delta $, respectively.}
\end{figure}

\end{document}